\documentclass[12pt]{article}
\catcode`\@=11
\def\vereq#1#2{\lower3pt\vbox{\baselineskip1.5pt \lineskip1.5pt
  \ialign{$\m@th#1\hfill##\hfil$\crcr#2\crcr\sim\crcr}}}
\def\gtrsim{\mathrel{\mathpalette\vereq>}}

\catcode`@=12

\def\lqcd{\Lambda_{\rm QCD}}
\def\FD{{\cal F}_D}
\def\FDs{{\cal F}_{D^*}}
\def\FDt{{\cal F}_{D^{(*)}}}
\begin{document}

\title{$|V_{cb}|$ and $|V_{ub}|$ from $B$ Decays: \\ 
  Recent Progress and Limitations\,\thanks{Invited talk at the Chicago 
  Conference on Kaon Physics (Kaon'99), June 21--26, 1999, Chicago, IL. 
  [FERMILAB-Conf-99/213-T] }}

\author{Zoltan Ligeti \\
  {\normalsize\it Theory Group, Fermilab, P.O. Box 500, Batavia, IL 60510}}

\date{}
\maketitle

\begin{abstract}

The determination of $|V_{cb}|$ and $|V_{ub}|$ from semileptonic $B$ decay is
reviewed with a critical discussion of the theoretical uncertainties.  Future
prospects and limitations are also discussed.

\end{abstract}

\section{Introduction}

The purpose of $K$ and $B$ physics in the near future is testing the
Cabibbo--Kobayashi--Maskawa (CKM) picture of quark mixing and $CP$ violation. 
The goal is to overconstrain the unitarity triangle by directly measuring the
sides and (some) angles in several decay modes.  If the value of $\sin2\beta$,
the $CP$ asymmetry in $B\to J/\psi\, K_S$, is near the CDF central
value~\cite{sin2beta}, then searching for new physics will require a
combination of precision measurements.  This talk concentrates on $|V_{cb}|$
and $|V_{ub}|$; the latter is particularly important since it largely controls
the experimentally allowed range for $\sin2\beta$ in the standard model.

\section{Exclusive decays}

In mesons composed of a heavy quark and a light antiquark (plus gluons and
$q\bar q$ pairs), the energy scale of strong processes is small compared to
the heavy quark mass.  The heavy quark acts as a static point-like color source
with fixed four-velocity, since the soft gluons responsible for confinement
cannot resolve structures much smaller than $\lqcd$, such as the heavy quark's
Compton wavelength.  Thus the configuration of the light degrees of freedom
become insensitive to the spin and flavor (mass) of the heavy quark, resulting
in a $SU(2n)$ spin-flavor symmetry~\cite{HQS} ($n$ is the number of heavy quark
flavors).  Heavy quark symmetry (HQS) helps understanding the spectroscopy and
decays of heavy hadrons from first principles.

The predictions of HQS are particularly restrictive for $\bar B\to
D^{(*)}\ell\bar\nu$ decays.  In the infinite mass limit all form factors are
proportional to a universal Isgur-Wise function, $\xi(v\cdot v')$, satisfying
$\xi(1)=1$~\cite{HQS}.  The symmetry breaking corrections can be organized in a
simultaneous expansion in $\alpha_s$ and $\lqcd/m_Q$ ($Q = c,b$).  The $\bar
B\to D^{(*)}\ell\bar\nu$ decay rates are given by
\begin{eqnarray}
{d\Gamma(\bar B\to D^*\ell\bar\nu)\over dw} &=& {G_F^2 m_B^5\over 48\pi^3}\, 
  r_*^3\, (1-r_*)^2\, \sqrt{w^2-1}\, (w+1)^2 \nonumber\\*
&& \times \left[ 1 + {4w\over 1+w} {1-2wr_*+r_*^2\over (1-r_*)^2} \right]
  |V_{cb}|^2\, \FDs^2(w) \,, \nonumber\\
{d\Gamma(\bar B\to D \ell\bar\nu)\over dw} &=& {G_F^2 m_B^5\over 48\pi^3}\, 
  r^3\, (1+r)^2\, (w^2-1)^{3/2}\, |V_{cb}|^2\, \FD^2(w) \,, 
\end{eqnarray}
where $w = v\cdot v'$ and $r_{(*)} = m_{D^{(*)}}/m_B$.  $\FDt(w)$ is equal to 
the Isgur-Wise function in the $m_Q \to\infty$ limit, and in particular 
$\FDt(1) = 1$, allowing for a model independent determination of $|V_{cb}|$.  
Including symmetry breaking corrections one finds
\begin{eqnarray}
\FDs(1) &=& 1 + c_A(\alpha_s) + {0\over m_Q} 
  + {(\ldots)\over m_Q^2} + \ldots \,, \nonumber\\*
\FD(1) &=& 1 + c_V(\alpha_s) + {(\ldots)\over m_Q} 
  + {(\ldots)\over m_Q^2} + \ldots \,.
\end{eqnarray}
The perturbative corrections, $c_A = -0.04$ and $c_V = 0.02$, have been
computed to order $\alpha_s^2$~\cite{Czar}, and the unknown higher order
corrections should affect $|V_{cb}|$ at below the 1\% level.  The vanishing of
the order $1/m_Q$ corrections to $\FDs(1)$ is known as Luke's
theorem~\cite{Luke}.  The terms indicated by $(\ldots)$ are only known using
phenomenological models at present.  Thus the determination of $|V_{cb}|$ from
$\bar B\to D^* \ell\bar\nu$ is theoretically more reliable than that from $\bar
B\to D\ell\bar\nu$ (unless using lattice QCD for $\FDt(1)$ --- see below),
although for example QCD sum rules predict that the order $1/m_Q$ correction to
$\FD(1)$ is small~\cite{LNN}.  Due to the extra $w^2-1$ suppression near zero
recoil, $\bar B\to D\ell\bar\nu$ is also harder experimentally.

The main uncertainty in this determination of $|V_{cb}|$ comes from the
estimate of nonperturbative corrections at zero recoil.  In the case of $\bar
B\to D^*\ell\bar\nu$, model calculations~\cite{BDsmodels} and sum rule
estimates~\cite{BDsumrules} suggest about $-5\%$.  Assigning a 100\% 
uncertainty to this estimate, I will use
\begin{equation}\label{F(1)}
\FDs(1) = 0.91\pm0.05 \,, \qquad \FD(1) = 1.02\pm0.08 \,.
\end{equation}
The most promising way to reduce these uncertainties may be calculating
directly the deviation of the form factor from unity, $\FDt(1)-1$, in lattice
QCD from certain double ratios of correlation functions~\cite{FNAL}.  Recent
quenched calculations give $\FD(1) = 1.06 \pm 0.02$ and $\FDs(1) = 0.935 \pm
0.03$~\cite{FNAL}, in agreement with Eq.~(\ref{F(1)}) but with smaller errors. 

Another uncertainty comes from extrapolating the experimentally measured
quantity, $|V_{cb}|\, \FDt(w)$, to zero recoil.  Recent theoretical
developments largely reduce this uncertainty by establishing a model
independent relationship between the slope and curvature of
$\FDt(w)$~\cite{Ben}.  This may also become less of an experimental problem at
asymmetric $B$ factories, where the efficiency may fall less rapidly near zero
recoil.

Eq.~(\ref{F(1)}) and the experimental average, $|V_{cb}| \FDs(1) = 0.0347
\pm 0.0015$~\cite{Stone}, obtained using the constraints on the shape of 
$\FDs(w)$ yield
\begin{equation}\label{Vcbexcl}
  |V_{cb}| = (38.1 \pm 1.7_{\rm exp} \pm 2.0_{\rm th}) \times 10^{-3} \,.  
\end{equation}
The value obtained from $\bar B\to D\ell\bar\nu$ is consistent with this, but
the experimental uncertainties are significantly larger.

For the determination of $|V_{ub}|$ from exclusive heavy to light decays, heavy
quark symmetry is less predictive.  It neither reduces the number of form
factors parameterizing these decays and nor determines the value of any form
factor.  Still, there are model independent relations between $B$ and $D$ decay
form factors, e.g., the form factors which occur in $D\to K^*\bar\ell\nu$ can
be related to those in $\bar B\to\rho\ell\bar\nu$ using heavy quark and chiral
symmetry~\cite{IsWi}.  These relations apply for the same value of $v\cdot v'$
in the two processes, i.e., from the measured $D\to K^*\bar\ell\nu$ form
factors one can predict the $\bar B\to \rho \ell\bar\nu$ rate in the large
$q^2$ region~\cite{lwold}.  Such a prediction has first order heavy quark and
chiral symmetry breaking corrections, each of which can be $15-20\%$.  Lattice
QCD also works best for large $q^2$, but the existing calculations are still
all quenched.  Light cone sum rules~\cite{Ball} are claimed to yield
predictions for the form factors with small model dependence in the small $q^2$
region.  Recently CLEO made the first attempt at concentrating at the large
$q^2$ region to reduce the model dependence, and obtained~\cite{CLEOBrho}
\begin{equation}
  |V_{ub}| = (3.25 \pm 0.14 ^{+0.21}_{-0.29} \pm 0.55) \times 10^{-3} \,.
\end{equation}

A determination of $|V_{ub}|$ from $\bar B\to\pi\ell\bar\nu$ is more
complicated because very near zero recoil ``pole contributions"~\cite{Bpi}
spoil the simple scaling of the form factors with the heavy quark mass.  Still,
in the future some combination of the soft pion limit, model independent bounds
based on dispersion relations and analyticity~\cite{Ben2}, and lattice results
may provide a determination of $|V_{ub}|$ from this decay with small errors.

If experimental data on the $D\to\rho\bar\ell\nu$ and $\bar B\to
K^*\ell\bar\ell$ form factors become available in the future, then $|V_{ub}|$
can be extracted with $\sim$10\% theoretical uncertainty~\cite{lwold} using a
``Grinstein-type double ratio"~\cite{Gtdr}, which only deviates from unity due
to corrections which violate both heavy quark and chiral symmetries.  Such a
determination is possible even if only the $q^2$ spectrum in
$D\to\rho\bar\ell\nu$ and the integrated $\bar B\to K^*\ell\bar\ell$ rate in
the large $q^2$ region are measured~\cite{lsw}.

\section{Inclusive decays}

Inclusive $B$ decay rates can be computed model independently in a series in
$\lqcd/m_b$ and $\alpha_s(m_b)$, using an operator product expansion
(OPE)~\cite{CGG,incl,MaWi}.  The $m_b\to\infty$ limit is given by $b$ quark
decay, and for most quantities of interest it is known including the dominant
part of the order $\alpha_s^2$ corrections.  Observables which do not depend on
the four-momentum of the hadronic final state (e.g., total decay rate and
lepton spectra) receive no correction at order $\lqcd/m_b$ when written in
terms of $m_b$, whereas differential rates with respect to hadronic variables
(e.g., hadronic energy and invariant mass spectra) also depend on
$\bar\Lambda/m_b$, where $\bar\Lambda$ is the $m_B-m_b$ mass difference in the
$m_b\to\infty$ limit.  At order $\lqcd^2/m_b^2$, the corrections are
parameterized by two hadronic matrix elements, usually denoted by $\lambda_1$
and $\lambda_2$.  The value $\lambda_2 \simeq 0.12\, {\rm GeV}^2$ is known from
the $B^*-B$ mass splitting.  Corrections to the $m_b\to\infty$ limit are
expected to be under control in parts of the $b\to q$ phase space where several
hadronic final states are allowed (but not required) to contribute with
invariant masses satisfying $m_{X_q}^2 \gtrsim m_q^2 + {\rm (few\ times)}\lqcd
m_b$.

The major uncertainty in the predictions for such ``sufficiently inclusive"
observables is from the values of the quark masses and $\lambda_1$, or
equivalently, the values of $\bar\Lambda$ and $\lambda_1$.  These quantities
can be extracted, for example, from heavy meson decay spectra.  A theoretical
subtlety is related to the fact that $\bar\Lambda$ (or the heavy quark pole
mass) cannot be defined unambiguously beyond perturbation
theory~\cite{renormalon}, and its value extracted from data using theoretical
expressions valid to different orders in the $\alpha_s$ may vary by order
$\lqcd$.  These ambiguities cancel~\cite{rencan} when one relates consistently
physical observables to one another.  One way to make this cancellation
manifest is by using short-distance quark mass definitions, but recent
determinations of such $b$ quark masses still have about $50-100\,$MeV
uncertainties~\cite{bmass}.

The shape of the lepton energy~\cite{gremmetal,Volo,GK} or hadronic invariant
mass~\cite{FLSmass1,FLSmass2,GK} spectra in $\bar B\to X_c \ell\bar\nu$ decay
can be used to determine $\bar\Lambda$ and $\lambda_1$.  Last year the CLEO
Collaboration measured the first two moments of the hadronic invariant
mass-squared distribution.  Each of these measurements gives an allowed band in
the $\bar\Lambda - \lambda_1$ plane, and their intersection
gives~\cite{CLEOparams}
\begin{equation}\label{cleonums}
\bar\Lambda = (0.33 \pm 0.08)\, {\rm GeV}\,, \qquad 
\lambda_1 = -(0.13 \pm 0.06)\,{\rm GeV}^2\,.  
\end{equation}
This result agrees well with the one obtained from an analysis of the lepton
energy spectrum in Ref.~\cite{gremmetal}.  CLEO also considered moments of the
lepton spectrum, however, without any restriction on the lepton energy,
yielding unlikely central values of $\bar\Lambda$ and $\lambda_1$.  Since this
analysis uses a model dependent extrapolation to $E_\ell < 0.6\,$GeV, I
consider the result in Eq.~(\ref{cleonums}) more reliable~\cite{dpf99}.  The
unknown order $\lqcd^3/m_b^3$ terms not included in Eq.~(\ref{cleonums})
introduce a sizable uncertainty~\cite{GK,FLSmass2}, which could be
significantly reduced when more precise data on the photon energy spectrum in
$\bar B\to X_s\gamma$ becomes available~\cite{LLMW,FLS}.  

The significance of Eq.~(\ref{cleonums}) is that, taken at face value, it gives
$|V_{cb}| = 0.0415$ from the $\bar B\to X_c \ell\bar\nu$ width with only 3\%
uncertainty.  The theoretical uncertainty hardest to quantify in the inclusive
determination of $|V_{cb}|$ is the size of quark-hadron duality
violation~\cite{Isgur}.  Studying the shapes of these $\bar B\to X_c
\ell\bar\nu$ decay distributions may be the best way to constrain this
experimentally, since it is unlikely that duality violation would not show up
in a comparison of moments of different spectra.  Thus, testing our
understanding of these spectra is important to assess the reliability of the
inclusive determination of $|V_{cb}|$, and especially that of $|V_{ub}|$ (see
below).

A new approach to replace the $b$ quark mass in theoretical predictions with
the $\Upsilon(1S)$ mass was proposed recently~\cite{upsexp}.  The crucial point
of this ``upsilon expansion" is that for theoretical consistency one must
combine different orders in the $\alpha_s$ perturbation series in the
expression for $B$ decay rates and $m_\Upsilon$ in terms of $m_b$.  As the
simplest example, consider schematically the $\bar B\to X_u \ell\bar\nu$ rate,
neglecting nonperturbative corrections,
\begin{equation}\label{bupole}
\Gamma(\bar B\to X_u \ell\bar\nu) = {G_F^2 |V_{ub}|^2\over 192\pi^3}\, m_b^5\,
  \bigg[ 1 - (\ldots) {\alpha_s\over\pi}\, \epsilon 
  - (\ldots) {\alpha_s^2\over\pi^2}\, \epsilon^2 - \ldots \bigg] \,. 
\end{equation}
The coefficients denoted by $(\ldots)$ are known, and the parameter $\epsilon
\equiv 1$ denotes the order in the upsilon expansion.  In comparison, the
expansion of the $\Upsilon(1S)$ mass in terms of $m_b$ has a different
structure,
\begin{equation}\label{upsmass}
m_\Upsilon = 2m_b \bigg[ 1 - (\ldots) {\alpha_s^2\over\pi^2}\, \epsilon
  - (\ldots) {\alpha_s^3\over\pi^3}\, \epsilon^2 - \ldots \bigg] \,,
\end{equation}
In this expansion one must assign to each term one less power of $\epsilon$
than the power of $\alpha_s$~\cite{upsexp}.  At the scale $\mu = m_b$ both of
these series appear badly behaved, but substituting Eq.~(\ref{upsmass}) into
Eq.~(\ref{bupole}) and collecting terms of a given order in $\epsilon$ 
gives~\cite{upsexp}
\begin{equation}\label{buups}
\Gamma(\bar B\to X_u \ell\bar\nu) = {G_F^2 |V_{ub}|^2\over 192\pi^3}\,
  \bigg({m_\Upsilon\over2}\bigg)^5\, \Big[ 1 - 0.115\epsilon
  - 0.035 \epsilon^2 - \ldots \Big] \,.
\end{equation}
The perturbation series, $1 - 0.115\epsilon - 0.035\epsilon^2$, is far better
behaved than the series in Eq.~(\ref{bupole}) in terms of the $b$ quark pole
mass, $1 - 0.17\epsilon - 0.13\epsilon^2$, or the series expressed in terms of
the $\overline{\rm MS}$ mass, $1+0.30\epsilon+0.19\epsilon^2$.  The uncertainty
in the decay rate using Eq.~(\ref{buups}) is much smaller than that in
Eq.~(\ref{bupole}), both because the perturbation series is better behaved, and
because $m_\Upsilon$ is better known (and better defined) than $m_b$.  The
relation between $|V_{ub}|$ and the $\bar B\to X_u \ell\bar\nu$ rate 
is~\cite{upsexp}
\begin{equation}\label{Vub}
|V_{ub}| = (3.06 \pm 0.08 \pm 0.08) \times 10^{-3}\,
  \bigg( {{\cal B}(\bar B\to X_u \ell\bar\nu)\over 0.001}
  {1.6\,{\rm ps}\over\tau_B} \bigg)^{1/2} \,.
\end{equation}
The upsilon expansion also improves the behavior of the perturbation series
for the $\bar B\to X_c \ell\bar\nu$ rate, and yields 
\begin{equation}\label{Vcb}
|V_{cb}| = (41.9 \pm 0.8 \pm 0.5 \pm 0.7) \times 10^{-3}\,
  \bigg( {{\cal B}(\bar B\to X_c \ell\bar\nu)\over0.105}\,
  {1.6\,{\rm ps}\over\tau_B} \bigg)^{1/2} \,.
\end{equation}
These results agree with other estimates~\cite{arnps} within the uncertainties.
The first error in Eqs.~(\ref{Vub}) and (\ref{Vcb}) come from assigning an
uncertainty equal to the size of the $\epsilon^2$ term, the second is from
assuming a $100\,$MeV uncertainty in Eq.~(\ref{upsmass}), and the third error
in Eq.~(\ref{Vcb}) is from a $0.25\,{\rm GeV}^2$ error in $\lambda_1$.  The
most important uncertainty is the size of nonperturbative contributions to
$m_\Upsilon$ other than those which can be absorbed into $m_b$, for which we
used $100\,$MeV.  By dimensional analysis it is of order $\lqcd^4/(m_b
\alpha_s)^3$, however, quantitative estimates vary in a large range.  It is
preferable to constrain such effects from data~\cite{LLMW,Andre}.

For the determination of $|V_{ub}|$, Eq.~(\ref{Vub}) is of little use by
itself, since ${\cal B}(\bar B\to X_u \ell\bar\nu)$ cannot be measured without
significant cuts on the phase space.  The traditional method for extracting
$|V_{ub}|$ involves a study of the electron energy spectrum in the endpoint
region $m_B/2 > E_\ell > (m_B^2-m_D^2)/2m_B$ (in the $B$ rest frame), which
must arise from $b \to u$ transition.  Since the width of this region is only
$300\,$MeV (of order $\lqcd$), an infinite set of terms in the OPE may be
important, and at the present time it is not known how to make a model
independent prediction for the spectrum in this region.  Another possibility
for extracting $|V_{ub}|$ is based on reconstructing the neutrino momentum. 
The idea is to infer the invariant mass-squared of the hadronic final state,
$s_H = (p_B - p_\ell - p_{\bar\nu})^2$.  Semileptonic $B$ decays satisfying
$s_H < m_D^2$ must come from $b \to u$ transition~\cite{FLW,DiUr,oldmass}.  

Both the invariant mass region $s_H < m_D^2$ and the electron endpoint region
$E_\ell > (m_B^2-m_D^2)/2m_B$ receive contributions from hadronic final states
with invariant masses between $m_\pi$ and $m_D$.  However, for the electron
endpoint region the contribution of states with masses nearer to $m_D$ is
strongly suppressed kinematically.  This region may be dominated by the $\pi$
and the $\rho$, and includes only of order 10\% of the total $\bar B\to X_u
\ell\bar\nu$ rate.  The situation is very different for the low invariant mass
region, $s_H < m_D^2$, where all such states contribute without any
preferential weighting towards the lowest mass ones.  In this case the $\pi$
and the $\rho$ exclusive modes comprise a smaller fraction, and only of order
10\% of the $\bar B\to X_u \ell\bar\nu$ rate is excluded from the $s_H < m_D$
region.  Consequently, it is much more likely that the first few terms in the
OPE provide an accurate description of the decay rate in the region $s_H <
m_D^2$ than in the region $E_\ell > (m_B^2-m_D^2)/2m_B$.  

Since $m_D^2$ is not much larger than $\lqcd m_b$, one needs to model the
nonperturbative effects in both cases.  However, assigning a 100\% uncertainty
to these estimates affects the extracted value of $|V_{ub}|$ much less from the
$s_H < m_D^2$ than from the $E_\ell > (m_B^2-m_D^2)/2m_B$ region.  Such
estimates suggest that the theoretical uncertainty in $|V_{ub}|$ determined
from the hadronic invariant mass spectrum in the region $s_H<m_D^2$ is about
$\sim$10\%.  If experimental constraints force to consider a significantly
smaller region, then the uncertainties increase rapidly.  The first analyses of
LEP data utilizing this idea were performed recently~\cite{Vubmass}, but it is
not transparent how they weigh the Dalitz plot, which affects crucially the
theoretical uncertainties.

The inclusive nonleptonic decay rate to ``wrong sign" charm ($\bar B\to
X_{u\bar c s}$) may also give a determination of $|V_{ub}|$ with modest
theoretical uncertainties~\cite{nonlept}, if such a measurement is
experimentally feasible.

\section{Conclusions}

The present status of $|V_{cb}|$ and $|V_{ub}|$ is approximately
\begin{equation}
|V_{cb}| = 0.040 \pm 0.002 \,, \qquad
|V_{ub} / V_{cb}| \simeq 0.090 \pm 0.025 \,.
\end{equation}
The central value and error of $|V_{cb}|$ comes from first principles, and the
uncertainty in both its exclusive and inclusive determination is of order
$1/m_Q^2$.  On the other hand, the above error on $|V_{ub}|$ is somewhat ad
hoc, since it is still estimated relying on phenomenological models.

Within the next 3--5 years, in my opinion, an optimistic scenario is roughly as
follows.  The theoretical error of $|V_{cb}|$ might be reduced to 2--3\%.  This
requires better agreement between the inclusive and exclusive determinations,
since in the exclusive determination the nonperturbative corrections to
$\FDt(1)$ are at the 5\% level and model dependent, while in the inclusive
determination it is hard to constrain model independently the size of
quark-hadron duality violation.  It will give confidence in lattice
calculations of $\FDs(1)$ and $\FD(1)$ if they give the same value of
$|V_{cb}|$, and the deviations of the form factor ratios conventionally denoted
by $R_{1,2}(w)$ from unity can also be predicted precisely.  Quark-hadron
duality violation in the inclusive determination of $|V_{cb}|$ can be
constrained by comparing the measured shapes of $\bar B\to X_c\ell\bar\nu$
decay spectra in different variables (e.g., lepton energy, hadronic invariant
mass, etc.).  

At the same time, the theoretical error of $|V_{ub}|$ might be reduced to about
10\%.  Again, a better agreement between the inclusive and exclusive
determinations is needed.  At this level only unquenched lattice calculations
will be trusted, and they ought to give consistent values of $|V_{ub}|$ from
\mbox{$\bar B\to\pi\ell\bar\nu$} and $\bar B\to\rho\ell\bar\nu$.  From
exclusive decays a double ratio method discussed in Sec.~2 may give $|V_{ub}|$
with $\sim$10\% error.  In inclusive $\bar B\to X_u\ell\bar\nu$ decay, the
hadron invariant mass spectrum should be measured up to a cut as close to $m_D$
as possible.  It would be reassuring as a check if varying this cut in some
range leaves $|V_{ub}|$ unaffected.

I would like to thank Jon Rosner and Bruce Winstein for inviting me and for 
organizing a very interesting and stimulating workshop.  
I also thank Adam Falk and Andreas Kronfeld for comments on the manuscript.
Fermilab is operated by Universities Research Association, Inc., under
DOE contract DE-AC02-76CH03000.

\end{document}